# Machine learning framework to predict the performance of lipid nanoparticles for nucleic acid delivery

*Gaurav Kumar* [†, *], *Arezoo M. Ardekani* [†]

[†] School of Mechanical Engineering, Purdue University, West Lafayette, Indiana, 47907, USA

[*]Email: kumar542@purdue.edu

**ABSTRACT**

Lipid nanoparticles (LNPs) are highly effective carriers for gene therapies, including mRNA and siRNA delivery, due to their ability to transport nucleic acids across biological membranes, low cytotoxicity, improved pharmacokinetics, and scalability. A typical approach to formulate LNPs is to establish a quantitative structure-activity relationship (QSAR) between their compositions and *in vitro*/*in vivo* activities which allows for the prediction of activity based on molecular structure. However, developing QSAR for LNPs can be challenging due to the complexity of multi-component formulations, interactions with biological membranes, and stability in physiological environments. To address these challenges, we developed a machine learning framework to predict the activity and cell viability of LNPs for nucleic acid delivery. We curated data from 6,398 LNP formulations in the literature, applied nine featurization techniques to extract chemical information, and trained five machine learning models for binary and multiclass classification. Our binary models achieved over 90% accuracy, while the multiclass models reached over 95% accuracy. Our results demonstrated that molecular descriptors, particularly when used with random forest and gradient boosting models, provided the most accurate predictions. Our findings also emphasized the need for large training datasets and comprehensive LNP composition details, such as constituent structures, molar ratios, nucleic acid types, and dosages, to enhance predictive performance.

## INTRODUCTION

Lipid nanoparticles (LNPs) have emerged across the pharmaceutical industry as promising vehicles to deliver a variety of therapeutic agents,[1–4] many of which are utilized in medical practice.[5–8] In particular, LNPs have shown exceptional promise as delivery systems for gene therapies, including mRNA and siRNA delivery.[9] These nanoparticles are engineered to encapsulate and transport nucleic acids, ensuring their safe and efficient transport into cells.[10] Owing to their size and properties, LNPs are taken up by cells *via* endocytosis. The ionizability of the lipids at low pH is believed to facilitate endosomal escape, releasing the cargo into the cytoplasm where mRNA can be translated into proteins or siRNA can induce gene silencing.[11] This technology has been pivotal in the rapid development of mRNA vaccines.[12,13] LNP-based mRNA vaccines are also in clinical trials for various infectious diseases, including Zika virus, tuberculosis, and influenza.[14]

LNPs designed for nucleic acid delivery typically consist of four key constituents: (1) ionizable lipids to encapsulate nucleic acids and facilitate endosomal escape, (2) helper lipids, and (3) cholesterol to enhance membrane rigidity, ensure particle stability, and improve intracellular delivery, and (4) a polyethylene glycol (PEG)-lipid conjugate to extend the circulation time of LNPs by preventing rapid clearance.[15,16] Seminal studies have shown that the chemical structures of LNP components can have a significant impact on transfection efficacy and organ selectivity.[16–19] For instance, Anderson and co-workers found that lipidoids with high delivery efficiency often possess common structural features, such as amide linkages, multiple alkyl tails of 8 to 12 carbons, and secondary amines.[20,21] Furthermore, the lipid composition can alter the structure and pKa of LNPs, affecting gene delivery efficiency to different cell types.[22] Optimizing the LNP composition, therefore, is essential to maximize gene delivery efficiency. Establishing a quantitative structure-activity relationship (QSAR) for LNPs is a common approach to achieve this. The elucidation of QSAR allows for the prediction of activity based on molecular structure, streamlining development, reducing empirical testing, and enabling tailored LNP formulations to meet specific therapeutic needs.[23,24]

Multiple studies have attempted to elucidate the structure-activity relationship by systematically varying the lipid structural variables, such as chain length, degree of saturation, and lipid/drug ratio.[18,25–28] For instance, Gonzalez-Diaz and co-workers used a perturbation approach to develop QSAR models, predicting a range of properties from ADME (absorption, distribution, metabolism, excretion) profiles to selecting components for cancer co-therapy drug-vitamin release nanoparticles and coated nanoparticles.[29–31] Metwally *et al.* applied a computer-assisted drug formulation design process to estimate the loading of different drugs on potential carriers.[32] However, investigating the structure-activity relationship of LNPs presents several significant challenges. Developing siRNA/mRNA LNPs is a lengthy and time-consuming process involving the chemical synthesis of numerous ionizable lipids and lipid-like molecules, followed by LNP formulation and subsequent *in vitro* and *in vivo* evaluations. The complexity and diversity of multi-component LNP formulations, their interactions with biological membranes, and their stability in various physiological environments hinder the establishment of clear QSAR models. Additionally, the lack of standardized protocols for LNP characterization and activity measurement can lead to variability and inconsistencies in data.[33]

To this end, machine learning (ML) models can handle large amounts of complex data and enable the understanding of the latent relationship between molecular features of LNPs and their biological functions. Before training ML algorithms, the constituent molecules of LNPs are transformed into numerical representations through a process called featurization. This process involves extracting and encoding various molecular attributes into numerical formats that capture the underlying chemical and physical properties of the molecules. There are several types of featurization techniques, each capturing different aspects of molecular properties. For instance, molecular descriptors quantify properties such as molecular weight, partial charge, partition coefficient, and the number of hydrogen bond donors. Fingerprints represent the presence or absence of specific substructures or patterns within a molecule using binary vectors. Molecular graphs treat molecules as graphs, where atoms are nodes and bonds are edges, allowing the use of graph-based algorithms and neural networks to capture complex relationships and interactions within the molecular structure.

In nanomedicine, several ML algorithms such as random forest (RF), gradient boosting (GB), support vector machine (SVM), and artificial neural networks (ANNs) have been employed to analyze structure-function relationships.[34–36] For instance, Le *et al.* used multiple linear regression, and Bayesian regularized ANNs on a dataset of 680 unique nanoparticles to predict their phase behavior.[37] Weissleder and co-workers used cluster analysis and SVM on a dataset of 109 unique nanoparticles to classify them based on activity and achieved a prediction accuracy of 73%.[38,39] Huang *et al.* used decision trees, SVM, and logistic regression (LR) on a dataset of 30 metal oxide nanoparticles, predicting their inflammatory potential with over 90% accuracy.[40] Other efforts have implemented various ML techniques for QSAR modeling of LNPs on datasets containing up to 325 unique LNPs.[41–43] However, given the vast chemical space of lipid compounds, models trained on such small datasets risk overfitting and may produce erroneous predictions when applied to a diverse set of LNPs. Additionally, most published studies omit molar ratios of LNP constituents and drug dosage during their model training.[38–41] The complex non-linear relationships between molecular features of lipids and their biological functions, interactions between different molecular features, and implementation of various learning algorithms pose additional challenges in consolidating learning from individual models trained on small datasets.

Over the last two decades, significant advances have been made in the combinatorial synthesis of ionizable lipids and LNPs, driven by the growing demand for efficient delivery systems in gene therapy and vaccine development. Researchers have employed high-throughput screening techniques to generate large libraries of lipid/lipid-like molecules, enabling the rapid identification of LNP candidates with optimal properties for encapsulating and delivering nucleic acids. For instance, 572 pH-switchable, multi-tailed ionizable phospholipids were synthesized *via* combinatorial reactions of amines with alkylated dioxaphospholane oxides.[18] The synthesis of ionizable lipids was followed by the synthesis of different LNP formulations encapsulating nucleic acids (such as Luciferase mRNA) by changing the ionizable lipid compound, or by varying the molar ratios of LNP constituents. The formulation of LNPs was followed by *in vitro* evaluations, and values for activity and cell viability were reported. The reported activity values, such as luciferase expression, indicate the amount of mRNA translated into proteins or the extent of siRNA-induced gene silencing, which represents the efficacy of nucleic acid delivery by LNPs. In this study, we curated data for 6398 LNP formulations from the literature. Using the curated data, we employed nine different featurization techniques, capture distinct aspects of molecular structures, to identify the most important molecular features for the accurate prediction of LNP performance. For the top hundred molecular features, we implemented several univariate and multivariate analysis techniques such as principal component analysis (PCA), t-distributed stochastic neighbor embedding (t-SNE), uniform manifold approximation and projection (UMAP), and K-means clustering (KMC) to explore the possibility of establishing QSAR using the molecular features of LNP constituents. Next, we developed a machine learning framework to predict the activity and cell viability of LNPs. Since the current study aims to predict the efficacy of LNPs for nucleic acid delivery and their cell viability, we will refer to the activity and cell viability of LNPs as our target variables. For each set of molecular features, we employed five different ML algorithms to identify the optimal learning algorithms for the accurate prediction of LNP performance. Different ML algorithms offer distinct benefits that can contribute to a comprehensive and robust analysis of LNP features. For instance, RF excels in handling high-dimensional data and provides insights into feature importance, making it useful for complex datasets with many variables. GB is highly effective in improving prediction accuracy by sequentially correcting errors made by previous models, making it particularly powerful for tasks requiring high precision. We used chemical features of constituent lipids, molar ratios of lipids, the ratio of lipids to nucleic acids, the type of encapsulated nucleic acids, and drug dosage as input features for the ML models. Our results, as discussed in the subsequent section, indicated that training on a large dataset of LNPs and including the LNP composition significantly improved the prediction accuracy.

**RESULTS AND DISCUSSION**

A bar plot showing the importance scores and a list of importance rank, feature ID, and feature name of the top hundred most important molecular features are presented in Figure S1 and Table S1, respectively, of the Supporting Information. Among the most important features were 'fr_unbrch_alkane', which refers to the number of unbranched alkanes of at least four members, Hall-Kier[44] 'Kappa3' and 'Kappa2', which reflect how close the molecule is to be cyclic or spherical, and the degree of branching in the molecule, maximum and minimum partial atomic charge in a molecule, van der Waals surface area (VSA) descriptors that measure the polarizability of a molecule, 'SlogP' that captures the hydrophobic and hydrophilic effects, and the electrotopological state (E-state). As shown in Figure 1, although there were observable differences in the mean feature value for LNPs with high and low activity, the large variability in feature values made it difficult to establish a clear distinction between the two classes. To quantitatively measure the distinction in feature values between LNPs with high and low activity, we calculated p-values from the t-test[45] and Mann-Whitney U-test,[46] and we calculated Cohen's D.[47] The t-test assesses whether the means of two classes (high vs. low) are statistically different from each other. The Mann-Whitney U-test is a non-parametric test that assesses whether two independent samples come from the same distribution. Cohen's D is a measure of effect size that quantifies the difference between two means relative to the variability in the data and is used to assess the practical significance of the difference between the feature values for the two classes. The calculation details for these metrics are presented in the Supporting Information. As shown in Table 1, p-values indicated a statistically significant difference between the feature values for LNPs with high and low activity. However, the value of Cohen's D was small (less than 0.5) for most features, indicating a practically insignificant difference between feature values for the two classes. This suggested that the features with high importance scores might capture subtle patterns or interactions within the data that were crucial for the model's overall performance. However, they may not be strong predictors of the class alone. A feature might be important because it interacts with other features or contributes to differentiating between classes in ways that are not immediately obvious from univariate analysis alone. To determine the interactions between the features, we computed the correlation matrix of the top hundred most important features, as shown in Figure S2 of the Supporting Information. We observed a strong correlation between multiple features (correlation coefficient above 0.7 or below -0.7) and a good correlation between most features (correlation coefficient above 0.5 or below -0.5). The high correlation coefficients suggested strong interactions between molecular features and explained why a univariate analysis, as shown in Figure 1, may be insufficient to distinguish between LNPs with high and low activity. Further multivariate analysis of molecular features, as shown in Figure S3 of the Supporting Information, could not reveal any clear distinction between LNPs with high and low activity. These analyses suggested complex interactions among molecular features and complex non-linear relationships between molecular features and target variables, necessitating the use of ML algorithms that can handle complex data.

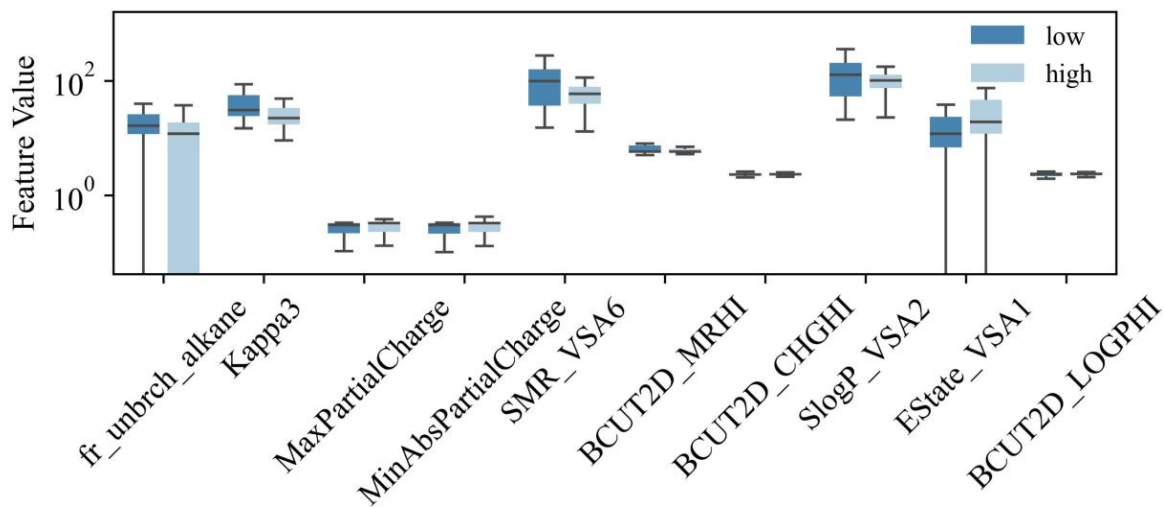

**Figure 1.** Box plots showing the feature values of the top ten most important features for LNPs with high and low activity. The feature values were computed using RDKit and represent specific molecular properties of each LNP.

**Table 1.** P-values for the t-test and Mann-Whitney U-test, along with Cohen's D, comparing feature values between LNPs with high and low activity.

| Feature name | t-test | U-test | Cohen's D |
|---|---|---|---|
| fr_unbrch_alkane | <0.001 | <0.001 | -0.57 |
| Kappa3 | <0.001 | <0.001 | -0.56 |
| MaxPartialCharge | <0.001 | <0.001 | 0.22 |
| MinAbsPartialCharge | 0.001 | <0.001 | 0.22 |
| SMR_VSA6 | <0.001 | <0.001 | -0.55 |
| BCUT2D_MRHI | <0.001 | 0.171 | -0.31 |
| BCUT2D_CHGHI | <0.001 | <0.001 | 0.28 |
| SlogP_VSA2 | <0.001 | <0.001 | -0.46 |
| EState_VSA1 | 0.191 | <0.001 | -0.09 |
| BCUT2D_LOGPHI | 0.012 | 0.224 | 0.17 |

Most ML models for predicting LNP properties in the existing literature have used relatively small training datasets, often comprising fewer than 260 LNP formulations. Furthermore, most studies omitted a comprehensive description of the LNP composition during their model training. The results in Figure 2 indicated that using small training sets could result in significantly lower prediction accuracy, precision, recall, and F1 scores. Specifically, models trained on 257 LNP formulations had average accuracy, precision, recall, and F1 scores of 0.43, 0.50, 0.43, and 0.46, respectively. Furthermore, all evaluation metrics showed considerable variability across different ML models and featurization techniques. Models trained on progressively larger datasets of LNP formulations (257, 480, 915, and 1875) showed noticeable improvements in the evaluation metrics, but the variability remained substantial. This implied that models trained on smaller datasets were not only less accurate in their predictions but also inconsistent, making them unreliable for practical applications where precise and reliable activity classification is crucial. In contrast, models trained on 3880 LNP formulations had average accuracy, precision, recall, and F1 scores of 0.85, 0.86, 0.91, and 0.89, respectively, with significantly less variability, indicating that the model's ability to make accurate and reliable predictions improved significantly as the training dataset size increased. Smaller datasets may not contain enough diverse and representative samples for the model to learn complex patterns, leading to overfitting and poor generalization to new data. The lower accuracy, precision, recall, and F1 scores for models trained on smaller datasets suggested that these models were likely capturing noise or spurious correlations rather than the true underlying relationships within the data. In contrast, the marked improvement in evaluation metrics for the model trained on 3880 formulations indicated that a larger dataset allowed the model to learn more effectively, capturing the true signal in the data and achieving better generalization. Therefore, these results highlight the importance of using sufficiently large and diverse datasets to train ML models, especially in complex domains like LNP classification, where capturing subtle differences in activity can be crucial for reliable predictions.

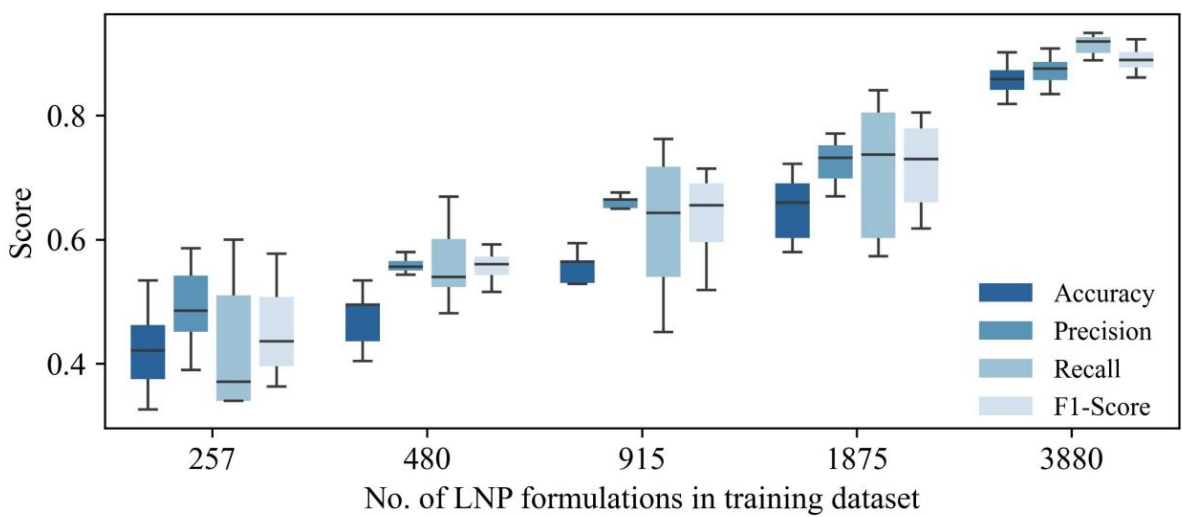

**Figure 2.** Accuracy, precision, recall, and F1 scores for binary classification of LNPs based on activity, plotted against training dataset size. Each plot highlights score variability across different featurization techniques and ML algorithms, as listed in Table 2.

As shown in Figure 3, when the input features included the chemical characteristics of all LNP constituents, their molar ratios, the type of encapsulated nucleic acids, the ratio of nucleic acids to lipids, and drug dosage, the models achieved significantly higher performance across all evaluation metrics. For instance, using input features generated using RDKit[48] descriptors with the RF algorithm, the accuracy of predictions improved from 82.6% (without composition) to 90.2% (with composition). Similarly, precision, recall, and F1 scores also showed marked improvements when the full compositional information was included. This improvement indicated that the activity of LNPs was likely influenced by multiple compositional factors, including the chemical structure and amount of all LNP constituents, the type and amount of nucleic acids, and the drug dosage.

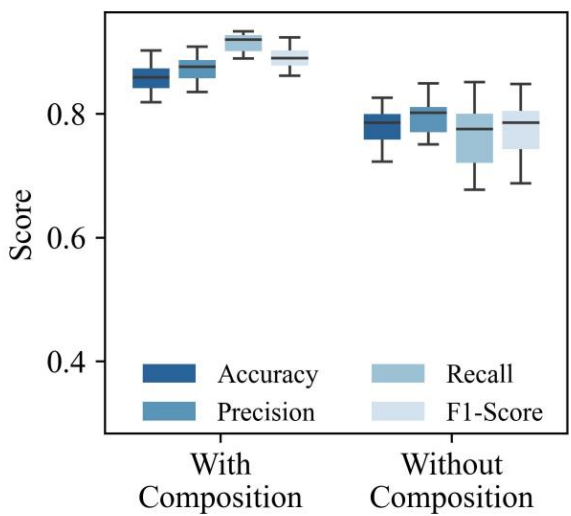

**Figure 3.** Comparison of ML model performance for binary classification of LNPs based on activity using two sets of input features: one with comprehensive composition information and one without. The 'with composition' feature set includes the chemical properties of all LNP constituents, their molar ratios, nucleic acid type, ratio of nucleic acids to lipids, and drug dosage. The 'without composition' set includes only the chemical properties of ionizable lipids.

For binary classification, as shown in Figure 4, our ML framework was able to distinguish between high- and low-activity LNPs with an accuracy of 90.21% and between high- and low-cell viability LNPs with an accuracy of 89.95%. For binary classification tasks, RF and GB algorithms performed substantially better than LR and linear SVM. For instance, with input features generated using RDKit descriptors, RF and GB were able to identify high-activity LNPs with accuracy scores of 90.21% and 89.18%, respectively, compared to accuracy scores of 79.28% and 84.43% for LR and linear SVM. The superior performance of RF and GB algorithms can be attributed to their ability to handle a large number of features while being able to capture the interactions among molecular features and complex non-linear relationships between the molecular features and activity of LNPs. Both RF and GB are also robust to redundant or irrelevant molecular features.

RF, in particular, is robust to noise in the data because it uses random subsets of molecular features at each split and aggregates the results from multiple trees, which tends to reduce the impact of noise or irrelevant features. Among various featurization techniques, descriptors yielded better results than fingerprints and molecular graphs in classifying LNPs based on activity. For instance, with RF as the ML model, features generated using RDKit descriptors and normalized descriptors yielded accuracy scores of 90.21% and 89.79%, respectively, compared to accuracy scores of 84.95% and 84.12% when RDKit fingerprints and PyTorch Geometric[49] were used. Molecular descriptors capture a wide range of physicochemical properties of molecules, and the corresponding high accuracy scores indicate that these descriptors are effective in representing the molecular features relevant to the activity of LNPs. Descriptors often include features, such as the number of tails in ionizable lipids, lipid molecule morphology, partial charge on atoms, and intramolecular electrostatic interactions, that are directly interpretable in terms of chemical or biological significance such as LNP stability, making them particularly useful for tasks where specific molecular properties are known to be important. In contrast, molecular fingerprints primarily capture structural information, namely, the presence or absence of specific substructures or functional groups, which are useful for tasks involving similarity searching. Similarly, molecular graphs capture local atomic environments and connectivity patterns, which are useful for modeling properties dependent on local structures. The superior performance of molecular descriptors over fingerprints and graphs suggests that a wide range of molecular properties beyond structural patterns and local features are crucial for the prediction of LNP activity. For the classification of LNPs based on cell viability, on the other hand, Extended Connectivity Fingerprints (ECFPs) performed the best with an accuracy score of 89.95%. The superior performance of fingerprints indicates the significance of substructures present in LNP constituents in predicting the cell viability of LNPs.

Similar to accuracy, RF and GB algorithms achieved substantially better precision and F1 scores than LR and linear SVM, as shown in Figure 5 and Figure 6. For binary classification of LNPs based on activity, the highest precision and F1 scores of 0.92 and 0.92, respectively, were achieved using RF on input features generated using RDKit descriptors, compared to the precision and F1 scores of 0.73 and 0.82, respectively, achieved using linear SVM on input features generated using PyTorch Geometric. In contrast, for binary classification of LNPs based on cell viability, the highest precision and F1 scores of 0.96 and 0.93, respectively, were achieved using GB on input features generated using ECFPs. High recall scores were also achieved for binary classification tasks, as shown in Figure S4 of the Supporting Information.

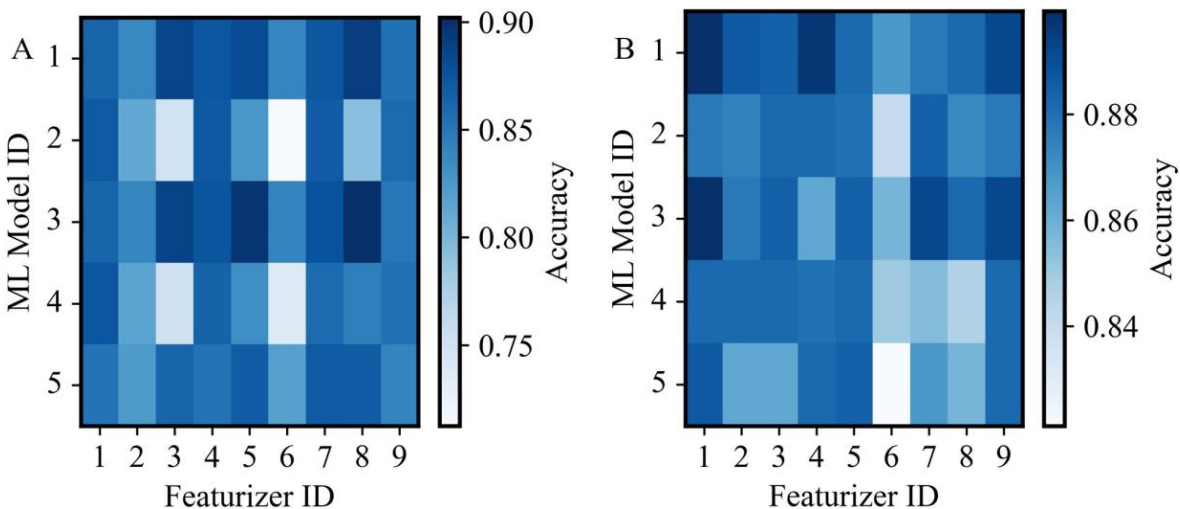

**Figure 4**. Accuracy scores for binary classification of LNPs into high or low class based on (A) activity and (B) cell viability, using various featurization techniques and ML algorithms, as listed in Table 2.

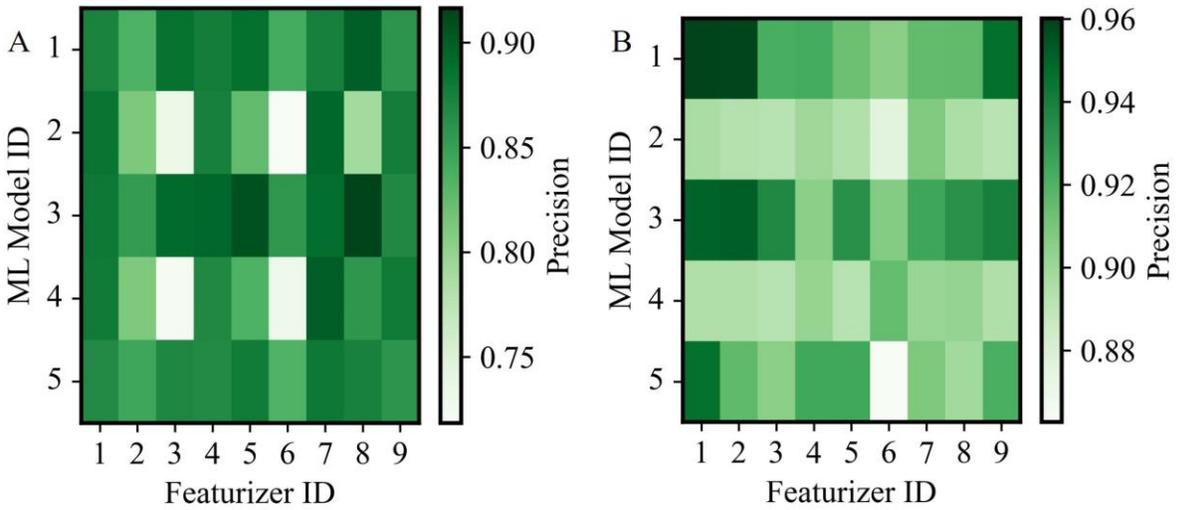

**Figure 5.** Precision scores for binary classification of LNPs into high or low class based on (A) activity and (B) cell viability, using various featurization techniques and ML algorithms, as listed in Table 2.

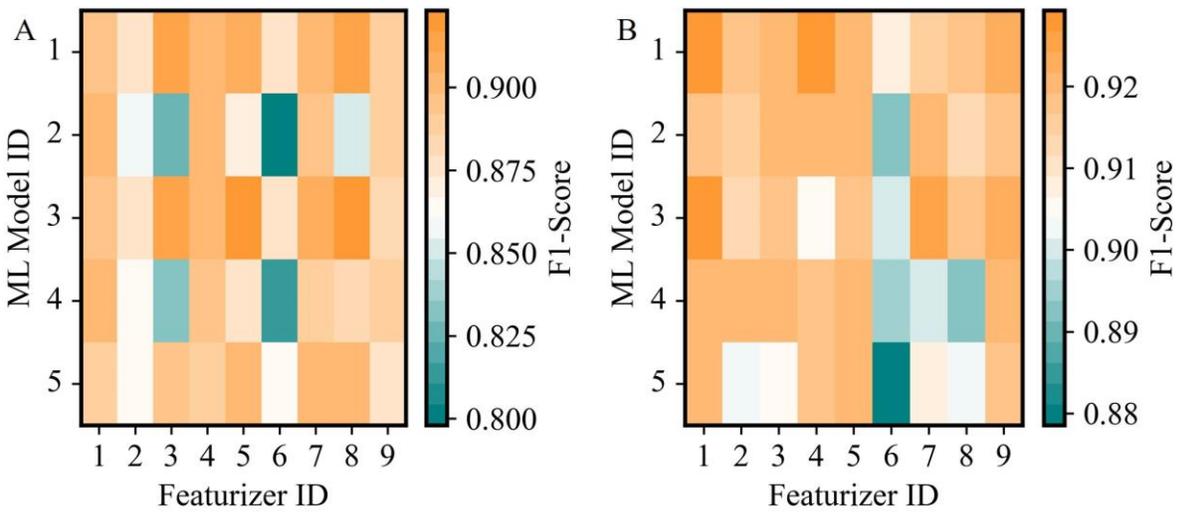

**Figure 6.** F1 scores for binary classification of LNPs into high or low class based on (A) activity and (B) cell viability, using various featurization techniques and ML algorithms, as listed in Table 2.

For multiclass classification of LNPs based on activity, our framework was able to identify high-activity LNPs with an accuracy of 95.15%, as shown in Figure 7A. Similar to binary classification, RF and GB algorithms with input features generated using descriptors yielded the highest accuracy scores. As shown in Figure 7B and C, we observed slightly lower precision and recall scores (~0.76), which pertain to false positive and false negative predictions, respectively. Lower scores for precision and recall suggested class imbalance in the dataset. Indeed, 34.03%, 30.19%, and 29.20% of the LNPs in the dataset were labeled as low, low-mid, and mid-high activity, respectively, whereas only 6.58% of the LNPs were labeled as high activity. As only a small percentage of all LNPs that were tested *in vitro* were reported to have high activity, a proportionately small percentage of LNPs were labeled as high activity in the training and testing datasets. Due to the presence of class imbalance, the ML models might be learning patterns that are specific to the majority class while not generalizing well to the minority class.

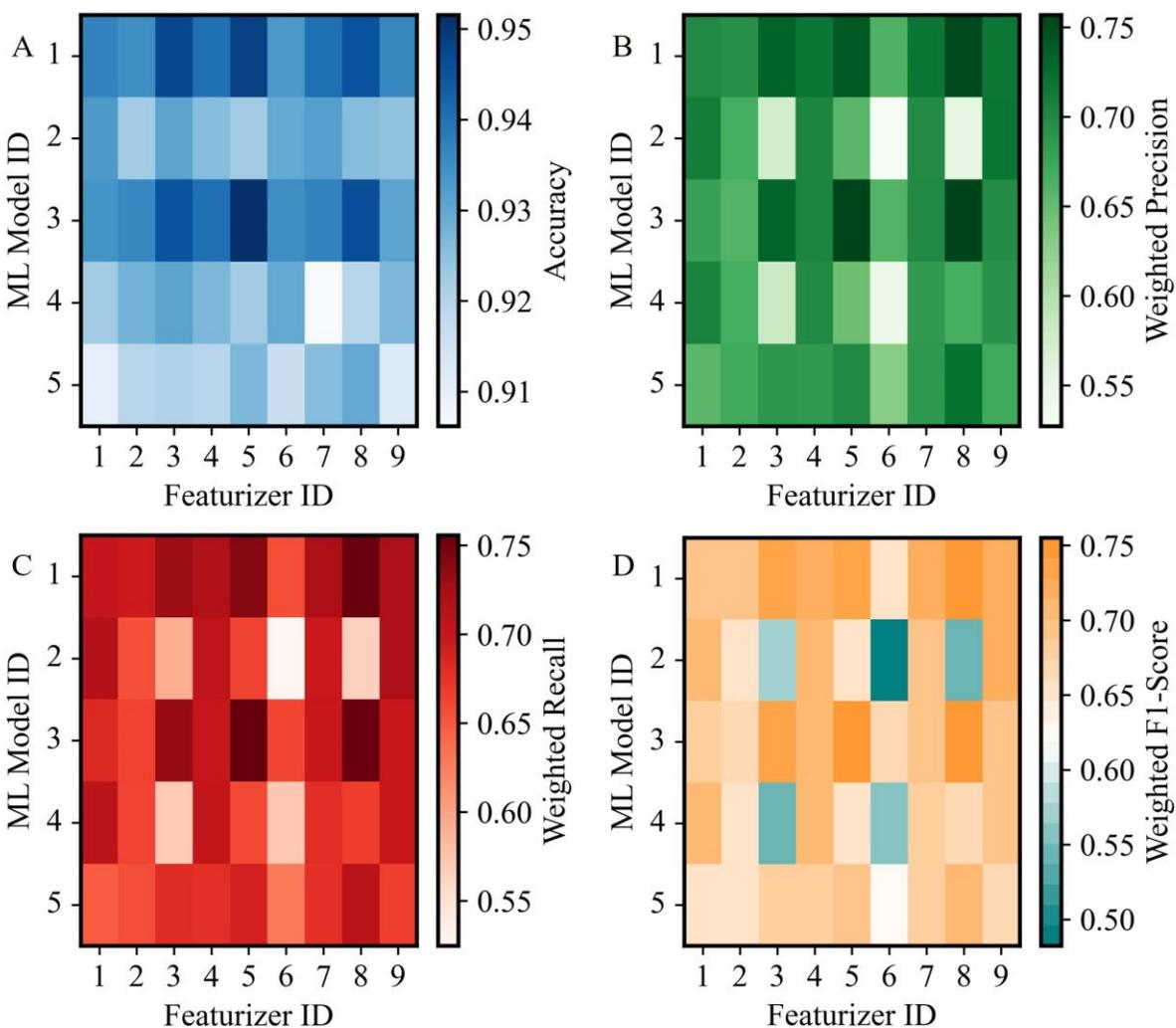

**Figure 7.** (A) Accuracy scores for identifying LNPs with high activity, (B) weighted precision, (C) weighted recall, and (D) weighted F1 scores for multiclass classification of LNPs based on activity, using various featurization techniques and ML algorithms, as listed in Table 2.

Besides weighted precision and recall scores, metrics such as Matthew's correlation coefficient (MCC)[50] and Cohen's kappa[51] scores can serve as balanced measures of the ML model performance for multiclass classification tasks with imbalanced datasets. MCC measures true and false positives and negatives across all classes, and a value of +1 indicates perfect prediction, 0 indicates random prediction, and -1 indicates total disagreement between prediction and true labels. Cohen's kappa measures the agreement between predicted and actual class labels, adjusting for the agreement that could occur by chance. A kappa value of +1 indicates perfect agreement, 0 indicates agreement no better than chance, and -1 indicates complete disagreement. As shown in Figure 8, with input features generated using descriptors, RF and GB algorithms achieved MCC and Cohen's kappa scores of ~0.65. An MCC score of 0.65 indicated a moderate to strong correlation between the predicted and true activity of LNPs and that the model was making informed decisions in correctly classifying LNPs with high, mid-high, low-mid, and low activity. A Cohen's kappa score of 0.65 indicated a substantial agreement between the true activity and the predicted activity of LNPs and showed that the model's predictions were consistent and aligned with the true distribution of classes, making it a reliable classifier, especially in a context where the cost of misclassification varies across classes. A high accuracy, combined with strong MCC and Cohen's Kappa scores, suggested that the model was both accurate in its predictions and reliable across different classes. The model was not only achieving a high overall rate of correct predictions but also performing well on both frequent and less frequent classes, even in the presence of any class imbalance. These results indicated that our model was likely generalizing well to new, unseen data and was capable of handling complex decision boundaries between multiple classes.

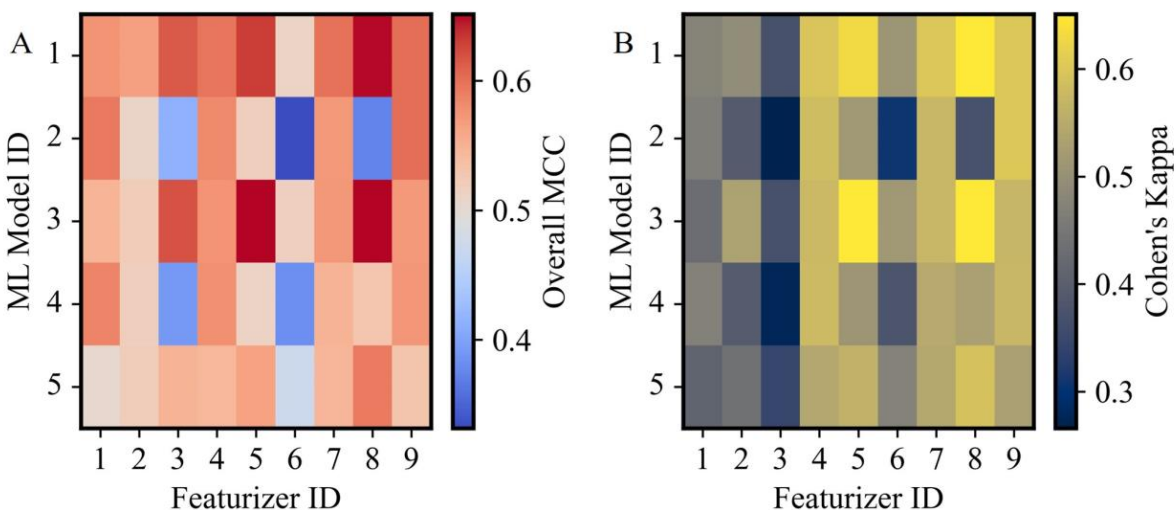

**Figure 8.** (A) Matthew's correlation coefficient and (B) Cohen's kappa score for multiclass classification of LNPs based on activity, using various featurization techniques and ML algorithms, as listed in Table 2.

The synthetic minority oversampling technique (SMOTE) is one of the widely used approaches to address class imbalance in a dataset by artificially generating new samples of the minority class. This is achieved by selecting two or more similar instances within the minority class and generating a new sample that is a combination of their features. By doing so, SMOTE increases the diversity of the minority class samples, reducing overfitting that could occur with simple replication and allowing the model to better learn the decision boundaries between the classes. We implemented SMOTE to improve the class balance in our dataset by oversampling the minority class (high activity). However, as shown in Figure S5 of the Supporting Information, the use of SMOTE resulted in marginal improvements in precision and recall scores. The poor performance of SMOTE was likely due to overlapping feature distributions of the high-activity class with other classes, as observed from the univariate and multivariate analyses. During SMOTE, if the decision boundaries between the high-activity class and other classes are highly non-linear or complex, the synthetic samples may not adequately represent the minority class, leading to poor learning by the model. Further investigation with more advanced techniques such as adaptive synthetic sampling, borderline-SMOTE, balanced random forest, and XGBoost with class weights may be more effective in addressing the issue of class imbalance.

While the machine learning models developed in this study have demonstrated strong predictive performance for the activity and cell viability of LNPs based on the chemical composition of their constituents, further investigations are needed to enhance the models' accuracy and applicability, especially for *in vivo* predictions. In this work, we generated input features primarily from the molecular descriptors, fingerprints, and graphs of the LNP constituent molecules, but incorporating additional characteristics of the LNPs themselves, such as size, polydispersity index, and zeta potential, could provide more comprehensive information that may lead to improved predictive power. These physical properties play a critical role in LNP behavior and efficacy, particularly in complex biological environments. Additionally, while we encoded the type of nucleic acids (0 for mRNA and 1 for siRNA), a more detailed representation that captures the structural and sequence-specific properties of the nucleic acids could further refine the model's predictions and is essential for *in vivo* applications where the interaction between nucleic acids and biological components is more complex. Moreover, the inclusion of features related to the target tissues or organs would be necessary for *in vivo* predictions, as the biodistribution and efficacy of LNPs are significantly influenced by their interactions with specific biological environments. Therefore, future studies should explore these additional dimensions to build more robust models that can accurately predict LNP performance across diverse biological settings.

**CONCLUSIONS**

In this study, we developed a machine learning framework to predict the activity and cell viability of lipid nanoparticles for nucleic acid delivery. We curated data for 6,398 LNPs from published literature, employed nine different featurization techniques to extract the chemical information of the LNP constituent molecules, and trained five different machine learning models for both binary and multiclass classification of LNPs based on their activity and cell viability. Our binary classification models achieved over 90% accuracy, while the multiclass classification models reached an accuracy of over 95%. The results demonstrated that using molecular descriptors, which capture a broad range of physicochemical properties, along with machine learning models such as random forest and gradient boosting, known for handling large, complex datasets and effectively capturing interactions among molecular features, produced the most accurate predictions. Additionally, we showed that models trained on smaller datasets were prone to inaccuracies and inconsistencies, highlighting the importance of a large training dataset for reliable and robust predictions. Our findings also indicated that including comprehensive LNP composition details, such as the chemical structure of all constituents, their molar ratios, the type of encapsulated nucleic acids, the ratio of nucleic acids to lipids, and drug dosage, substantially enhances the predictive accuracy of the models.

**METHODS**

Our machine learning framework, a schematic of which is shown in Figure 9, consisted of four phases: (1) Data preparation, which included data curation, data processing, and null model testing; (2) Feature extraction, which involved transforming the chemical structure of LNPs into numerical representations that can be used as input features for machine learning models; (3) Machine learning which included splitting the dataset for training, testing and hyperparameter tuning, followed by model training and generating predictions for the target variables; and (4) Evaluation of predictions followed by subsequent analysis.

The process of data curation involved collecting data for the chemical structure and molar ratio of LNP constituents, type of encapsulated nucleic acids (mRNA or siRNA), ratio of lipid to nucleic acids, drug dosage, and values of activity and cell viability from *in vitro* evaluations. As shown in Table S2 of the Supporting Information, we utilized 16 different studies to collect data for a total of 6398 LNP formulations. For each LNP, we used ChemDraw v22.2.0 to draw the chemical structure of the constituent molecules and obtained their simplified molecular-input line-entry system (SMILES) strings using RDKit. We used WebPlotDigitizer[52] and manual data curation on published figures to get the values of the target variables for each LNP. In different studies, the target variables were reported either as numeric values, such as a luciferase expression value of 50,000 relative light units (RLUs), or as a range of values, such as between 10,000 and 100,000 RLUs. Due to differences in experimental conditions, measurement techniques, and reporting formats across different studies, the values of target variables collected from different studies must be normalized to a common scale before training machine learning models to ensure consistency and comparability across diverse datasets. For each study, if the target variables were reported as numeric values, we applied a log-transformation to the target variables which helped to stabilize the variance across different levels of the target variable. Next, we normalized the log-transformed values using a min-max normalization as $t_{j,i}^n = \frac{t_{j,i} - t_{i,min}}{t_{i,max} - t_{i,min}}$ where the index $i$ represents a dataset ($i$= 1 to 16), $t_{j,i}$ is the value of the target variable for $j^{th}$ LNP in dataset $i$, $t_{i,min}$ is the lowest value of the target variable in dataset $i$, $t_{i,max}$ is the highest value of the target variable in dataset $i$, and $t_{j,i}^n$ is the normalized value of the target variable for LNP $j$ in dataset $i$. This min-max normalization ensured that the value of the target variable in each dataset ranged between 0 and 1. We further transformed the normalized values of the target variables into categorical labels such as "high" and "low" for a binary classification and "high", "mid-high", "low-mid" and "low" for a multiclass classification. This transformation helped to mitigate the noise and variability inherent in continuous measurements, allowing the ML models to focus on distinct classes that represent biologically, or experimentally significant thresholds. It also simplified the complexity of continuous data, making it more manageable and interpretable for classification algorithms. To transform the normalized target variables into categorical labels, we used Scikit-learn's[53] KMC algorithm to identify two and four clusters of data for binary and multiclass classification, respectively. For binary classification, LNPs within the cluster with higher values were labeled as "high" and those within the cluster with lower values were labeled as "low". Similarly, for multiclass classification, LNPs within the cluster with the highest values were labeled as "high", followed by "mid-high", "low-mid", and "low". The details of the KMC algorithm are presented in the Supporting Information. For studies where the target variables were reported in ranges, the range with the highest value was labeled as "high", and the remaining ranges were labeled as "low" for binary classification. For multiclass classification, the top three highest value ranges were labeled as "high", "mid-high", and "low-mid", respectively, and the remaining ranges were labeled as "low". For the activity, we generated two sets of labels: one for binary and one for multiclass classification. For cell viability, we generated one set of labels for binary classification only.

After processing the data, we used three null models to establish a benchmark for evaluating the performance of our ML models that would be subsequently trained. If a complex ML model does not perform significantly better than a null model, it suggests that the ML model may be overfitting or not capturing the underlying structure of the data. If the ML model performs well on training data but not much better than the null model on test data, it indicates overfitting. Null models can also reveal challenges in the data, such as class imbalance or noise. We used a Random Guessing null model, which predicts classes of the target variables randomly based on their distribution in the training data, a Majority Class Predictor null model, which always predicts the most frequent class, and a Stratified Random Guessing null model, which instead of randomly guessing any class, predicts the classes according to their frequencies in the training data while maintaining the overall distribution of classes in the predictions. Further details of the null models and their performance are presented in Table S3 and Table S4 of the Supporting Information.

To prepare the curated data for ML algorithms, the chemical composition of LNPs was transformed into numerical representations through featurization. The first step in featurization involved extracting and encoding various molecular attributes of LNP constituents into numerical formats. Since different featurization techniques capture distinct aspects of molecular structures, which can significantly influence the performance of ML models, we employed nine different molecular featurizers, as shown in Table 2, to extract physicochemical features of all LNP constituents, including ionizable lipids, helper lipids, cholesterol, and PEG-lipids. A detailed description of each featurization technique is provided in the Supporting Information. For each LNP constituent, a featurizer generated a vector of numerical values. We then calculated the weighted features for each constituent by multiplying its feature vector by its molar ratio. Subsequently, for each LNP, we computed a feature vector by summing the weighted features of its constituents. Finally, we added three additional entries to the feature vector: one representing the ratio of nucleic acids to lipids, one indicating the type of nucleic acid (0 for mRNA and 1 for siRNA), and one representing the drug dosage in nanograms. This final step resulted in the complete feature vector for each LNP. The above steps were repeated for all nine featurizers listed in Table 2, producing nine distinct sets of input features.

For 209 molecular features generated using RDKit Descriptors, we calculated the relative importance of these features based on the mean decrease in impurity[53], also known as Gini importance. The importance scores are relative measures, namely, features with higher importance scores are more influential in determining the predictions made by the model. They are the features that the model relies on most to make accurate predictions. However, high importance does not necessarily mean that the feature alone determines the outcome. It often reflects that the feature interacts with other features in meaningful ways, leading to more accurate splits in the decision trees. Features with low or zero importance might either be irrelevant for the prediction task or redundant, namely, the information they provide is already captured by other features. Using the top hundred most important features, we performed several univariate and multivariate analyses, such as PCA, t-SNE, UMAP, and KMC, to explore any data clusters or trends that could facilitate the distinction between LNPs with high and low performance. The details of the above analysis techniques are provided in the Supporting Information.

We used five different ML algorithms, as shown in Table 2, for binary and multiclass classification of the target variables (activity and cell viability). The machine learning phase involved data splitting, hyperparameter tuning, model training, and classification. We used 1,549 LNP formulations (Data14, 15, and 16 as shown in Table S1 of Supporting Information) for hyperparameter tuning of various ML algorithms. The details of hyperparameter tuning are shown in the Supporting Information. For the remaining 4,849 LNP formulations, 80 percent of the data was used for model training, and 20 percent of the data was used for model testing. After data splitting and hyperparameter tuning, we trained 5 ML algorithms on the training datasets, each using 9 different sets of input features, resulting in $5 \times 9 \times 2 = 90$ trained models for activity and $5 \times 9 = 45$ trained models for cell viability. Finally, each model was used to predict the class of all LNPs in the testing dataset for the corresponding target variables.

After generating predictions, the performance of the ML models was evaluated using several metrics for binary and multiclass classification tasks. For binary classifiers, accuracy, precision, recall, and F1 score were employed as the primary evaluation criteria. Accuracy measures the proportion of correctly classified instances out of the total instances, providing a general sense of the model's effectiveness. However, accuracy alone can be misleading, especially in cases of imbalanced datasets, where the number of instances in one class significantly outweighs the other. Therefore, additional metrics such as precision, recall, and F1 score were utilized to provide a more nuanced evaluation of the model's performance. Precision quantifies the proportion of true positive predictions among all instances predicted as positive, indicating how many of the predicted positive instances are actually positive. Recall measures the proportion of true positive predictions among all actual positive instances, reflecting the model's ability to capture all relevant instances. The F1 score is the harmonic mean of precision and recall, providing a balanced metric that considers both false positives and false negatives. For multiclass classification, the aforementioned metrics were adapted and computed for each class in a one-vs-rest scheme. In addition, we calculated two additional metrics: MCC, and Cohen's kappa to evaluate the performance of multiclass classifiers. The calculation details of the evaluation metrics are presented in the Supporting Information.

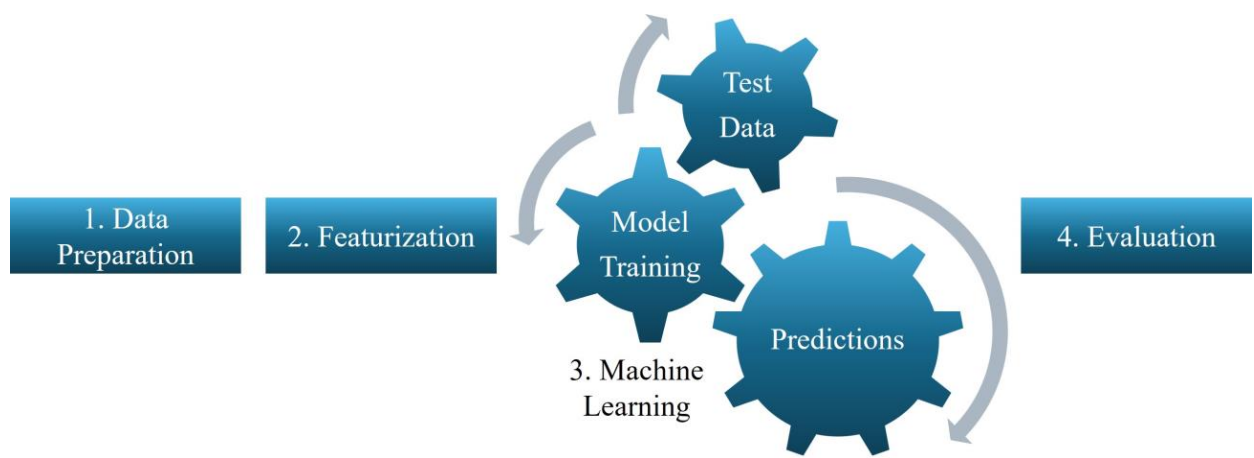

**Figure 9.** Schematic representation of the machine learning framework.

**Table 2.** List of featurization techniques and machine learning algorithms implemented in this study.

| Featurizer ID | Featurizer name | ML model ID | ML model name |
| --- | --- | --- | --- |
| 1 | Extended connectivity fingerprints | 1 | Gradient boosting |
| 2 | MACCS keys | 2 | Logistic regression |
| 3 | PAGTN graphs | 3 | Random forest |
| 4 | Morgan counts | 4 | Linear support vector machine |
| 5 | Normalized descriptors | 5 | K nearest neighbors |
| 6 | PyTorch Geometric | | |
| 7 | RDKit atom pairs | | |
| 8 | RDKit descriptors | | |
| 9 | RDKit fingerprints | | |

## ASSOCIATED CONTENT

### Supporting Information

Plots showing the hundred most important features and their correlation matrix, and calculation details for t-test, Mann-Whitney U-test and Cohen's D; results of multivariate analysis techniques employed on the top hundred important features; recall scores for binary classification of LNPs based on activity and cell viability; precision and recall scores for multiclass classification of LNPs based on activity after implementing SMOTE; list of published literature that was used for data curation; description of the K-means clustering algorithm that was used for labeling target variables; description of three null models and their performance; description of nine featurization techniques that were used to generate input features for all LNPs; description of metrics used for evaluation of binary and multiclass classifiers; results of hyperparameter tuning.

## AUTHOR INFORMATION

### Corresponding author

Gaurav Kumar – School of Mechanical Engineering, Purdue University, West Lafayette, Indiana, 47907, USA

Email: kumar542@purdue.edu

### Author

Arezoo M. Ardekani – School of Mechanical Engineering, Purdue University, West Lafayette, Indiana, 47907, USA

Email: ardekani@purdue.edu

**ACKNOWLEDGMENT**

This work was supported by Eli Lilly and Company.

# Supporting Information for

# Machine learning framework to predict the performance of lipid nanoparticles for nucleic acid delivery

*Gaurav Kumar [†, *], Arezoo M. Ardekani [†]*

[†] School of Mechanical Engineering, Purdue University, West Lafayette, Indiana, 47907, USA

[*]Email: kumar542@purdue.edu

**S1. Feature Importance Analysis**

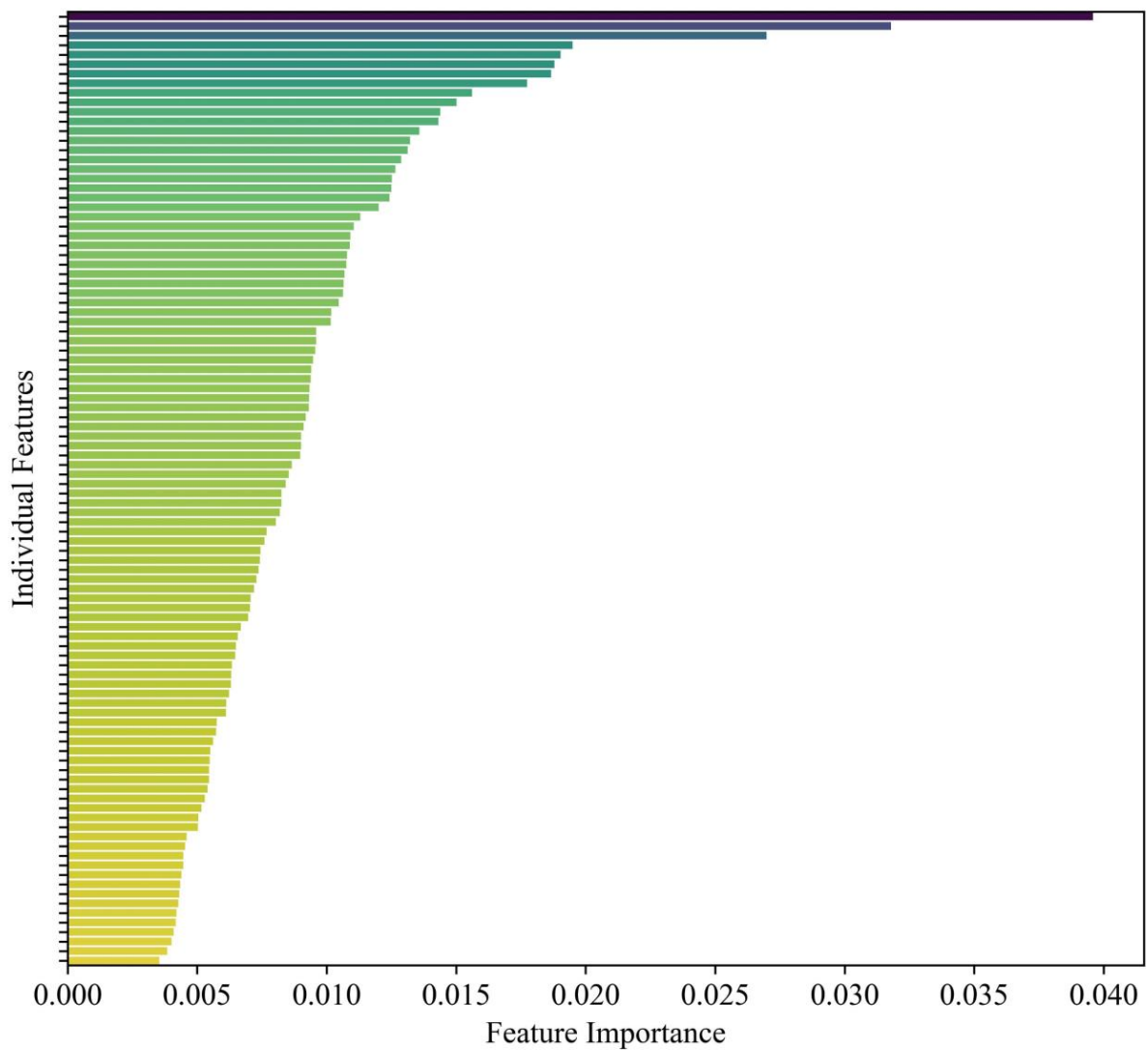

**Figure S10**. Feature importance scores for the top hundred features generated using RDKit descriptors. The y-axis bars represent individual features ranked by their importance scores, as detailed in Table S3.

**Table S3.** List of the top hundred most important features generated using RDKit descriptors. Feature importance scores were calculated based on mean decrease in impurity.

| Importance rank | Feature ID | Feature name |
| --- | --- | --- |
| 1 | 208 | fr_unbrch_alkane |
| 2 | 45 | Kappa3 |
| 3 | 12 | MaxPartialCharge |
| 4 | 15 | MinAbsPartialCharge |
| 5 | 67 | SMR_VSA6 |
| 6 | 25 | BCUT2D_MRHI |
| 7 | 21 | BCUT2D_CHGHI |
| 8 | 75 | SlogP_VSA2 |
| 9 | 84 | EState_VSA1 |
| 10 | 23 | BCUT2D_LOGPHI |
| 11 | 29 | BertzCT |
| 12 | 44 | Kappa2 |
| 13 | 32 | Chi0v |
| 14 | 24 | BCUT2D_LOGPLOW |
| 15 | 88 | EState_VSA3 |
| 16 | 105 | FractionCSP3 |
| 17 | 5 | qed |
| 18 | 4 | MinEStateIndex |
| 19 | 22 | BCUT2D_CHGLO |
| 20 | 1 | MaxAbsEStateIndex |
| 21 | 2 | MaxEStateIndex |
| 22 | 58 | PEOE_VSA7 |
| 23 | 98 | VSA_EState3 |
| 24 | 100 | VSA_EState5 |
| 25 | 97 | VSA_EState2 |
| 26 | 30 | Chi0 |
| 27 | 89 | EState_VSA4 |
| 28 | 103 | VSA_EState8 |
| 29 | 8 | HeavyAtomMolWt |
| 30 | 35 | Chi1v |
| 31 | 7 | MolWt |
| 32 | 99 | VSA_EState4 |
| 33 | 46 | LabuteASA |
| 34 | 9 | ExactMolWt |
| 35 | 102 | VSA_EState7 |
| 36 | 10 | NumValenceElectrons |
| 37 | 34 | Chi1n |
| 38 | 68 | SMR_VSA7 |
| 39 | 106 | HeavyAtomCount |
| 40 | 31 | Chi0n |
| 41 | 27 | AvgIpc |
| 42 | 49 | PEOE_VSA11 |
| 43 | 28 | BalabanJ |
| 44 | 124 | MolMR |
| 45 | 85 | EState_VSA10 |
| 46 | 3 | MinAbsEStateIndex |
| 47 | 95 | VSA_EState1 |
| 48 | 79 | SlogP_VSA6 |
| 49 | 43 | Kappa1 |
| 50 | 37 | Chi2v |
| 51 | 33 | Chi1 |
| 52 | 39 | Chi3v |
| 53 | 26 | BCUT2D_MRLOW |
| 54 | 41 | Chi4v |
| 55 | 40 | Chi4n |
| 56 | 42 | HallKierAlpha |
| 57 | 19 | BCUT2D_MWHI |
| 58 | 6 | SPS |
| 59 | 140 | fr_NH0 |
| 60 | 16 | FpDensityMorgan1 |
| 61 | 78 | SlogP_VSA5 |
| 62 | 74 | SlogP_VSA12 |
| 63 | 20 | BCUT2D_MWLOW |
| 64 | 123 | MolLogP |
| 65 | 66 | SMR_VSA5 |
| 66 | 54 | PEOE_VSA3 |
| 67 | 116 | NumHDonors |
| 68 | 17 | FpDensityMorgan2 |
| 69 | 38 | Chi3n |
| 70 | 115 | NumHAcceptors |
| 71 | 127 | fr_Al_OH_noTert |
| 72 | 104 | VSA_EState9 |
| 73 | 18 | FpDensityMorgan3 |
| 74 | 126 | fr_Al_OH |
| 75 | 47 | PEOE_VSA1 |
| 76 | 61 | SMR_VSA1 |
| 77 | 93 | EState_VSA8 |
| 78 | 59 | PEOE_VSA8 |
| 79 | 76 | SlogP_VSA3 |
| 80 | 118 | NumRotatableBonds |
| 81 | 57 | PEOE_VSA6 |
| 82 | 91 | EState_VSA6 |
| 83 | 36 | Chi2n |
| 84 | 83 | TPSA |
| 85 | 62 | SMR_VSA10 |
| 86 | 107 | NHOHCount |
| 87 | 60 | PEOE_VSA9 |
| 88 | 64 | SMR_VSA3 |
| 89 | 14 | MaxAbsPartialCharge |
| 90 | 48 | PEOE_VSA10 |
| 91 | 94 | EState_VSA9 |
| 92 | 92 | EState_VSA7 |
| 93 | 87 | EState_VSA2 |
| 94 | 151 | fr_allylic_oxid |
| 95 | 101 | VSA_EState6 |
| 96 | 52 | PEOE_VSA14 |
| 97 | 71 | SlogP_VSA1 |
| 98 | 13 | MinPartialCharge |
| 99 | 90 | EState_VSA5 |
| 100 | 53 | PEOE_VSA2 |

To quantitatively measure the distinction in feature values between lipid nanoparticles (LNPs) with high and low activity, we calculated p-values from the t-test and Mann-Whitney U test, and we calculated Cohen's D. The t-test is a parametric test used to compare the means of two groups to determine if there is a statistically significant difference between them. It assumes normally distributed data with similar variances between groups. For a t-test comparing two groups (feature values for LNPs with high and low activity) with means $\bar{X}_1$ and $\bar{X}_2$, sample sizes $n_1$ and $n_2$, and standard deviations $s_1$ and $s_2$, the t-statistic is calculated as $t = \frac{\bar{X}_1 - \bar{X}_2}{\sqrt{\frac{s_1^2}{n_1} + \frac{s_2^2}{n_2}}}$. The **p-value** is calculated based on the t-statistic and the degrees of freedom ($df$), typically using a t-distribution table. The degrees of freedom for the independent t-test can be approximated as $df = \frac{\left(\frac{s_1^2}{n_1} + \frac{s_2^2}{n_2}\right)^2}{\frac{\left(\frac{s_1^2}{n_1}\right)^2}{n_1 - 1} + \frac{\left(\frac{s_2^2}{n_2}\right)^2}{n_2 - 1}}$. The Mann-Whitney U-test is a non-parametric test used to compare two independent groups when the data does not necessarily follow a normal distribution. It assesses whether one group tends to have larger or smaller values than the other. For two groups of sample sizes $n_1$ and $n_2$, and the sums of ranks $R_1$ and $R_2$, the U-statistic is the smaller of $U_1$ or $U_2$, where $U_1 = n_1 n_2 \frac{n_1(n_1+1)}{2} - R_1$ and $U_2 = n_1 n_2 \frac{n_2(n_2+1)}{2} - R_2$. The **p-value** is determined based on the U-statistic and the sample sizes of both groups. Cohen's D is a measure of effect size that quantifies the difference between two group means in terms of standard deviations. It is used to indicate the magnitude of the difference between groups, regardless of sample size. Cohen's D is calculated as $d = \frac{\bar{X}_1 - \bar{X}_2}{S_{pooled}}$, where $S_{pooled}$ is the pooled standard deviation calculated as $S_{pooled} = \sqrt{\frac{(n_1 - 1)s_1^2 + (n_2 - 1)s_2^2}{n_1 + n_2 - 2}}$.

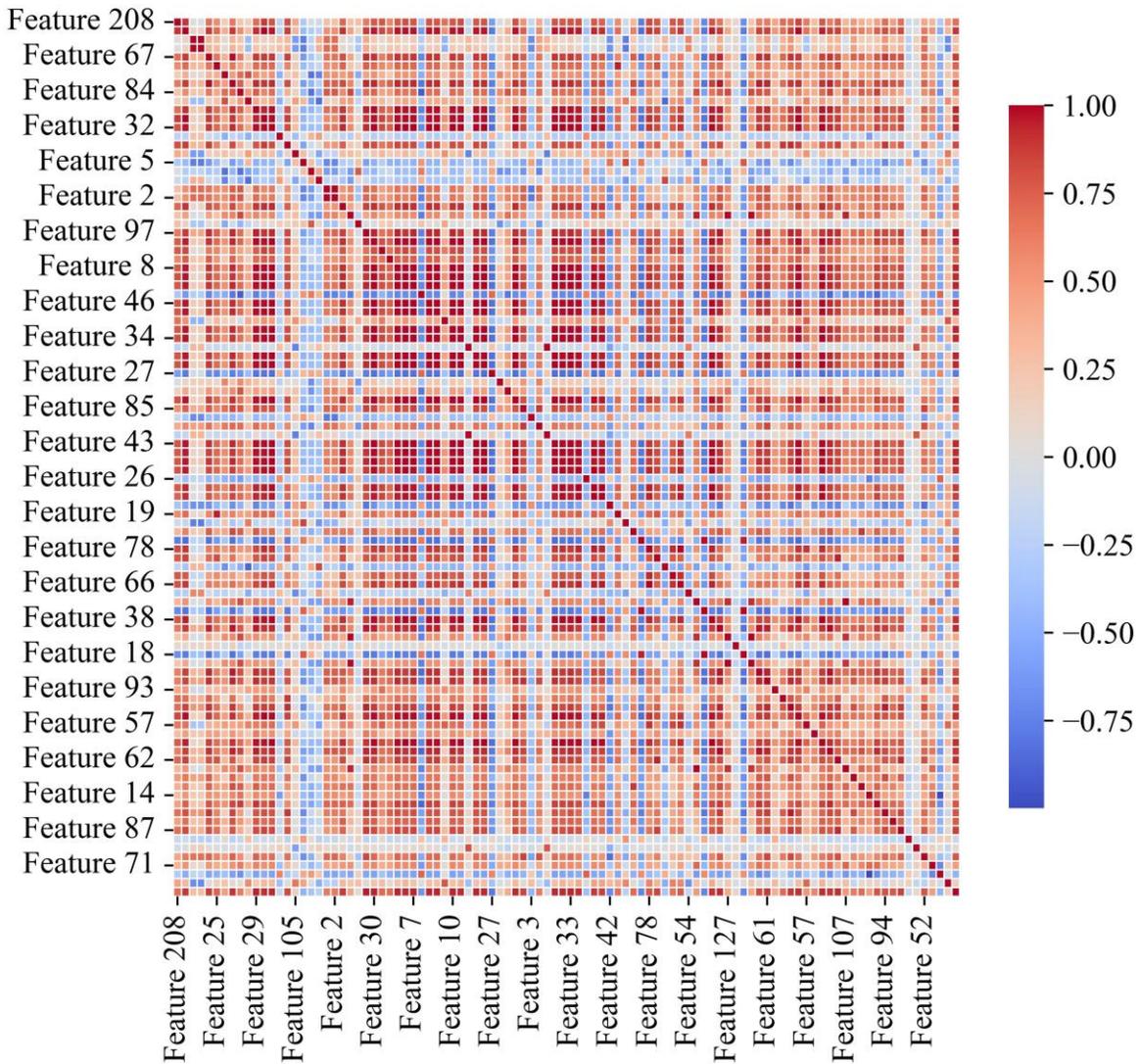

**Figure S11.** Pearson's correlation matrix for the top hundred most important features. A cool-warm colormap was used to represent correlation coefficient values.

## S2. Multivariate Analysis

**Principal component analysis (PCA)** is a dimensionality reduction technique used in machine learning to transform a large set of correlated variables into a smaller set of uncorrelated variables called principal components. The goal of PCA is to reduce the complexity of the data while retaining as much of the original variability as possible. PCA works by finding the directions, or axes, along which the data varies the most. These directions are known as the principal components. The first principal component captures the greatest variance in the data, the second principal component captures the next highest variance orthogonal to the first, and so on. To achieve this, PCA calculates the eigenvectors and eigenvalues of the data's covariance matrix, where the eigenvectors represent the directions of maximum variance (the principal components), and the eigenvalues indicate the amount of variance explained by each principal component. The data is then projected onto these principal components, resulting in a new set of linearly uncorrelated variables. This transformation reduces the dimensionality of the data while preserving its most significant patterns and structures, making it easier to visualize, analyze, and use for tasks such as classification, regression, or clustering. In the current study, we implemented PCA on the values of the top hundred most important features for all LNP formulations in the testing dataset. Following PCA, we plotted the data projected onto the first two principal components, as shown in Figure S12A.

**The t-distributed stochastic neighbor embedding (t-SNE)** is a nonlinear dimensionality reduction technique used for visualizing high-dimensional data in a lower-dimensional space. t-SNE is particularly effective at preserving the local structure of the data, meaning it maintains the relative distances between similar data points while projecting them onto a lower-dimensional space. The algorithm works by first converting the high-dimensional Euclidean distances between data points into conditional probabilities that represent pairwise similarities. It then defines a similar probability distribution in the lower-dimensional space and aims to minimize the difference between these two distributions using Kullback-Leibler divergence, a measure of how one probability distribution diverges from a second, expected probability distribution. To achieve this, t-SNE uses a gradient descent optimization technique. t-SNE uses t-distribution to model the probability distribution in the lower-dimensional space, which helps to prevent the crowding problem, where too many points get clustered together when reducing dimensions. In the current study, we used Scikit-learn to implement t-SNE on the values of the top hundred most important features to project the data onto two dimensions, as shown in Figure S12B.

**Uniform manifold approximation and projection (UMAP)** is a nonlinear dimensionality reduction technique that is particularly effective for visualizing high-dimensional data in a lower-dimensional space, typically two or three dimensions. UMAP is based on mathematical foundations from manifold theory and topological data analysis. The algorithm works by first constructing a high-dimensional graph representation of the data, where each data point is connected to its nearest neighbors, capturing the local structure of the data. UMAP then optimizes this graph to project the data into a lower-dimensional space while preserving the global and local structure as much as possible. Unlike other dimensionality reduction techniques like t-SNE, UMAP tends to preserve more of the global structure of the data, making it better suited for understanding both local clusters and broader patterns. In the current study, we used Scikit-learn to implement UMAP on the values of the top hundred most important features to project the data onto two dimensions, as shown in Figure S12C.

We also used the K-means clustering (KMC) algorithm on the values of the top hundred most important features to identify two clusters of data. We then visualized the two clusters of data, as shown in Figure S12D, to check if the two data clusters had a significant overlap with high- and low-activity LNPs.

The lack of clear distinction between LNPs with high and low activity after implementing the above multivariate analysis techniques suggested that the classes with high and low activity might be overlapping in the feature space, or that the relationships between these features and activity were too complex or non-linear to be captured by these methods. It also suggested the use of more sophisticated modeling techniques that can capture subtle, non-linear patterns in the data.

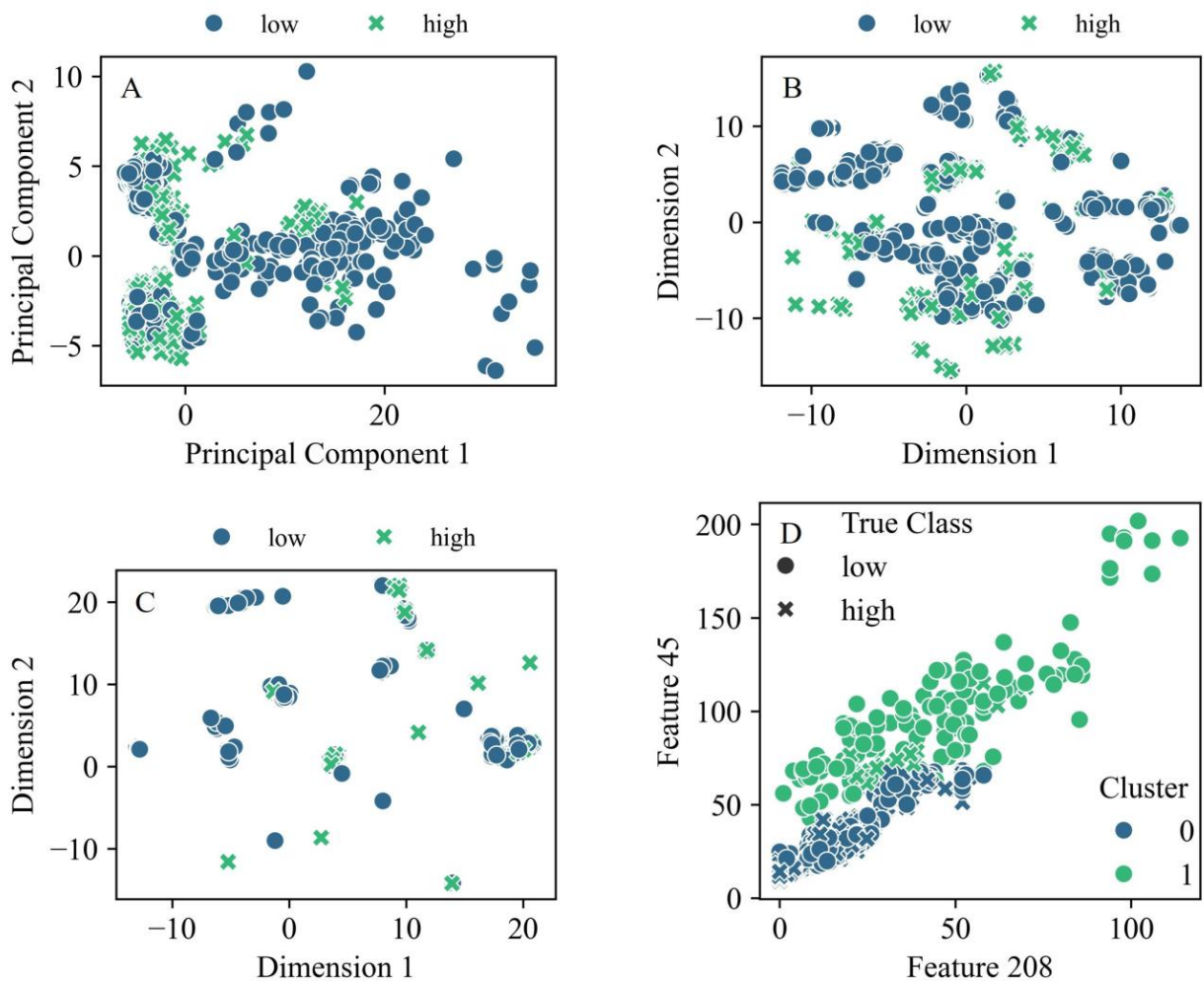

**Figure S12.** (A) Principal component analysis , (B) t-distributed stochastic neighbor embedding, (C) uniform manifold approximation and projection, and (D) K-means clustering applied to the top hundred most important molecular features.

**S3. Recall scores for binary classification task**

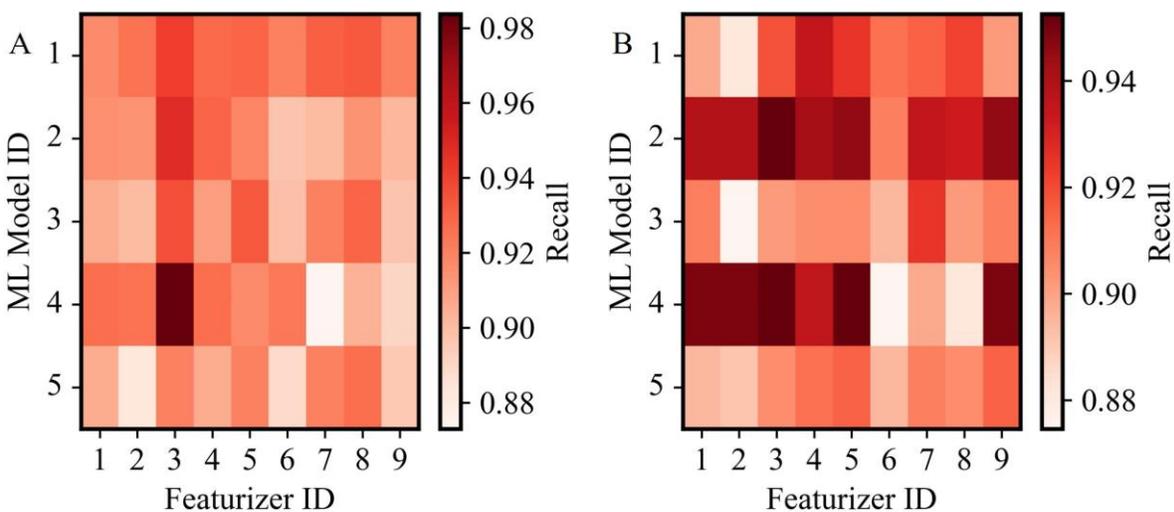

**Figure S13.** Recall scores for binary classification of LNPs into high or low class based on (A) activity and (B) cell viability, using various featurization techniques and machine learning (ML) algorithms, as listed in Table 1.

**S4. Evaluation after implementing synthetic minority oversampling technique (SMOTE)**

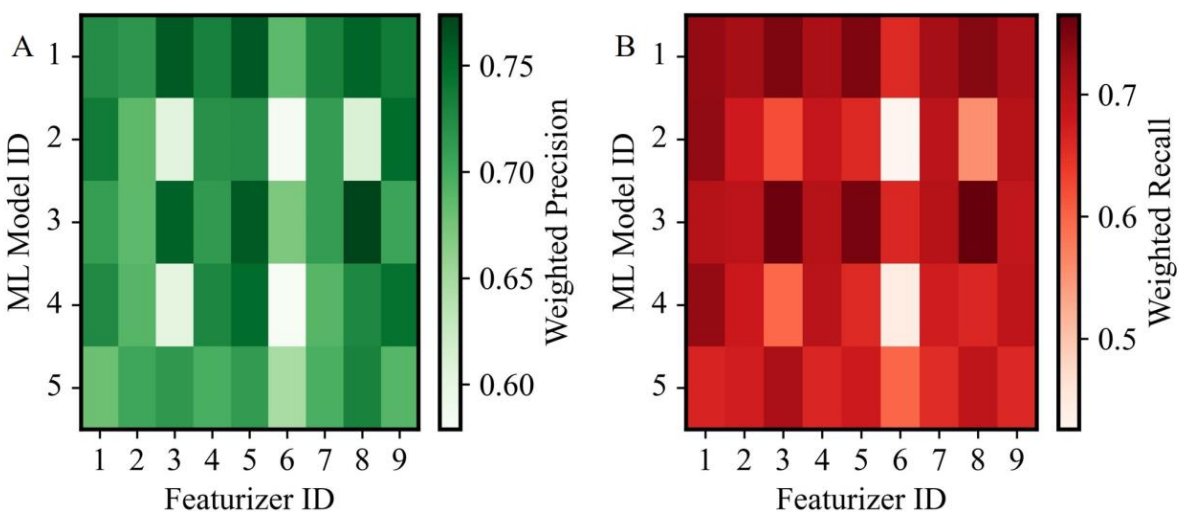

**Figure S14.** (A) Weighted precision and (B) weighted recall scores for multiclass classification of LNPs based on activity, after implementing SMOTE.

**S5. List of sources used for data curation**

**Table S4.** List of sources used for data curation, including the number of LNP formulations curated from each source. Data4 includes 104 unique formulations tested across different cell lines.

| Data ID | No. of LNP formulations |
|---|---|
| Data1[1] | 572 |
| Data2[2] | 91 |
| Data3[3] | 40 |
| Data4[4] | 257 |
| Data5[5] | 435 |
| Data6[6] | 480 |
| Data7[7] | 96 |
| Data8[8] | 78 |
| Data9[9] | 65 |
| Data10[10] | 35 |
| Data11[11] | 408 |
| Data12[12] | 60 |
| Data13[13] | 2232 |
| Data14[14] | 1512 |
| Data15[15] | 24 |
| Data16[16] | 13 |
| Total | 6398 |

## S6. K-means clustering for labeling target variables

KMC is an unsupervised machine learning algorithm that partitions a dataset into **K** distinct, non-overlapping groups or clusters, where each data point belongs to the cluster with the nearest mean. The algorithm begins by selecting **K** initial cluster centroids using k-means++, which selects initial cluster centroids using sampling based on an empirical probability distribution of the points' contribution to the overall inertia. The inertia is defined as the sum of squared distances of data points to their closest cluster center. The algorithm then assigns each data point to the nearest centroid based on the Euclidean distance, forming **K** clusters. Once all points are assigned, the centroids are recalculated as the mean of the data points in each cluster. This assignment and centroid recalculation process iterates until convergence, which is typically defined as a point where the centroids no longer change significantly, or the data point assignments remain stable between iterations.

For categorical labeling of our target variables, we used Scikit-learn's KMC algorithm using the following parameters: n_clusters= 2 for binary labeling and n_clusters= 4 for multiclass labeling, init='k-means++', and max_iter=300 which defines the maximum number of iterations.

### S7. Null models

A **random guessing null model** makes predictions purely by chance, without considering any patterns or relationships in the data. For a classification problem, the model randomly assigns each instance to one of the available classes, with each class having an equal probability of being chosen. This approach represents the lowest possible performance benchmark for any classifier, as it provides no meaningful information about the data. If a trained model performs no better than random guessing, it indicates that the model has failed to learn any significant patterns from the data. A **majority class predictor null model** always predicts the most frequent class in the training dataset, regardless of the input features. This approach is particularly relevant in the context of imbalanced datasets, where one class significantly outnumbers the others. The majority predictor model will have a high accuracy if the dataset is heavily skewed toward one class, but its performance on other metrics like precision, recall, or F1 score may be poor, especially for the minority classes. A **stratified random guessing null model** improves upon simple random guessing by considering the class distribution in the dataset. Instead of assigning each instance to a class with equal probability, this model randomly assigns instances to classes based on their relative frequencies in the dataset. As an example, if a dataset consists of 70% of instances in Class A and 30% in Class B, the stratified random guessing model will assign a class to a new instance with a 70% probability of selecting Class A and a 30% probability of selecting Class B. This approach provides a more informed baseline than pure random guessing, particularly in datasets with imbalanced class distributions. The performance of a stratified random guessing model is often used to set a benchmark that accounts for class imbalances when evaluating more advanced models. The performance of the above three null models for binary and multiclass classification of LNPs based on activity is shown in Table S5 and Table S6, respectively.

As shown in Table S5, for the binary classification task, the accuracy score is 0.5 for random guessing and 0.63 for the majority class predictor, indicating a slight imbalance in the dataset. For the multiclass classification task, as shown in Table S6, the accuracy of the random guessing model is close to 0.25, and the majority class represents ~34% of the dataset.

**Table S5.** Performance of null models in the binary classification of LNPs based on activity.

| Null Model | Accuracy | Precision | Recall | F1 score |
|---|---|---|---|---|
| Random Guessing | 0.5 | 0.63 | 0.5 | 0.56 |
| Majority Class Predictor | 0.63 | 0.63 | 1 | 0.77 |
| Stratified Random Guessing | 0.53 | 0.63 | 0.62 | 0.62 |
52

**Table S6.** Performance of null models in the multiclass classification of LNPs based on activity.

| Null Model | Accuracy | Precision | Recall | F1 score |
|---|---|---|---|---|
| Random Guessing | 0.24 | 0.33 | 0.23 | 0.27 |
| Majority Class Predictor | 0.34 | 0.34 | 1 | 0.51 |
| Stratified Random Guessing | 0.30 | 0.33 | 0.33 | 0.33 |

**S8. Featurization techniques**

**Extended Connectivity Fingerprints (ECFPs)**[17] are a type of molecular fingerprint that provides a detailed representation of a molecule's structural features by capturing the connectivity patterns around each atom in concentric layers. ECFPs are generated using a modified version of the Morgan algorithm, where each atom in a molecule is initially assigned a unique identifier based on its atomic properties (such as atomic number, hybridization, and the number of bonded hydrogens). The algorithm then iteratively updates these identifiers by expanding in concentric layers around each atom up to a specified radius, combining the identifiers of neighboring atoms to create unique substructure patterns. These patterns are then hashed into a fixed-length binary or count vector, which represents the presence or frequency of specific substructures within the molecule. ECFPs are particularly effective in applications like virtual screening, molecular similarity searches, and quantitative structure-activity relationship (QSAR) modeling because they capture both the local environment of individual atoms and the broader molecular topology. We used DeepChem[18] to generate ECFPs using the following parameters: radius = 2 which represents the fingerprint radius, size = 2048 which represents the length of the fingerprint vector, chiral=

False which ignores chirality in fingerprint generation, and bonds= True which considers bond order in fingerprint generation.

Molecular Access System (MACCS) keys are a predefined set of 166 structural keys that capture the presence or absence of specific chemical substructures within a molecule. Each key corresponds to a particular chemical property or structural feature, such as the presence of an aromatic ring, a specific atom type, or a certain bond pattern. When generating MACCS keys for a molecule, each key is set to "1" if the corresponding substructure is present and "0" if it is absent, resulting in a binary vector that succinctly captures the molecule's structural characteristics. Due to their simplicity and interpretability, MACCS keys are widely used in cheminformatics for various applications, including similarity searching, clustering, virtual screening, and machine learning tasks. While they may not capture the full complexity of a molecule's structure like some more advanced fingerprints, MACCS keys provide a quick and computationally efficient way to characterize molecules. We used DeepChem to generate MACCS keys fingerprints.

Path-Augmented Graph Transformer Network (PAGTN)[19] is a graph-based featurization technique that represents molecules as graphs where atoms are nodes and bonds are edges. This technique leverages graph neural networks (GNNs) to capture complex molecular structures and their relationships by learning from both local atomic environments and global molecular contexts. PAGTN incorporates augmented paths in the graph representation to improve the model's ability to capture long-range dependencies, making it effective for various predictive modeling tasks, such as molecular property prediction. We used DeepChem to generate PAGTN-based molecular graphs.

Morgan counts represent molecules using the count of specific substructures as defined by the Morgan algorithm. Unlike traditional binary Morgan fingerprints, which indicate the presence or absence of particular molecular substructures, Morgan counts provide a count-based representation. Each count corresponds to the frequency of specific substructures, or molecular fragments, within a molecule. The Morgan count algorithm starts by assigning a unique identifier to each atom in the molecule based on its atomic properties (such as atomic number and connectivity). It then iteratively expands around each atom up to a specified radius, generating unique identifiers that represent substructures of increasing complexity. Instead of merely noting whether a substructure exists, the Morgan counts capture how many times each substructure appears within the molecule. This count-based approach offers a more nuanced and detailed representation, making it particularly useful in applications such as QSAR modeling, molecular similarity searches, and clustering. We used RDKit to generate Morgan counts-based fingerprints.

PyTorch Geometric is a library that provides tools for deep learning on graph-structured data, including molecular graphs. It enables the construction and training of GNNs that can learn complex patterns from molecular graphs where atoms are nodes and bonds are edges. This framework is particularly well-suited for capturing both the local and global structural information in molecules, making it highly effective for various cheminformatics tasks, such as molecular property prediction, reaction outcome prediction, and molecular docking.

Atom pair fingerprints are a type of molecular descriptor that captures the relationship between pairs of atoms in a molecule along with the shortest path between them, including the types of bonds and the number of bonds in that path. It encodes each pair of atoms by combining the atom types (based on properties like atomic number, hybridization, aromaticity, etc.) and the distance (in terms of the number of bonds) between them, producing a binary vector representation of the molecule. This representation captures both local and extended structural information within a molecule, making it useful for tasks such as molecular similarity searches and clustering. We used RDKit to generate the atom pair fingerprint vectors of size 2048.

**RDKit descriptors** are a comprehensive set of molecular descriptors provided by RDKit, an open-source cheminformatics library. These descriptors quantify various physical, chemical, and structural properties of molecules, and they are widely used in computational chemistry, machine learning, and drug discovery. The descriptors include a broad range of features, such as molecular weight, logP (partition coefficient, indicative of lipophilicity), the number of rotatable bonds, hydrogen bond donors and acceptors, topological polar surface area (TPSA), and other topological, geometric, and electronic properties. In normalized descriptors, the RDKit descriptors are scaled/standardized to ensure they are on a similar range or scale, which helps prevent any single descriptor from disproportionately influencing machine learning models due to differences in numerical magnitude.

RDKit fingerprints are a type of molecular fingerprint that represents molecules using a fixed-length binary vector, where each bit indicates the presence or absence of a particular molecular substructure or pattern. They are commonly used in cheminformatics for tasks like similarity searches, clustering, and molecular property prediction. RDKit fingerprints are generated using hashing algorithms and can be customized based on various parameters, such as fingerprint length and path length, making them flexible and effective for a wide range of applications in molecular modeling and drug discovery.

**S9. Evaluation metrics for classification tasks**

To evaluate the performance of the binary classifier, we calculated the metrics of accuracy, precision, recall, and F1 score. **Accuracy** measures the proportion of correctly predicted instances (both true positives and true negatives) out of the total number of instances and provides a general sense of the model's performance. The accuracy score was calculated as $\frac{TP+TN}{TP+FP+TN+FN}$, where $TP$ is true positive, $FP$ is false positive, $TN$ is true negative, and $FN$ is false negative. **Precision** quantifies the ratio of correctly predicted positive instances (true positives) to the total predicted positives (true positives plus false positives), indicating how many of the predicted positive results are actually positive. The precision score was calculated as $\frac{TP}{TP+FP}$. **Recall** calculates the proportion of true positive predictions out of all actual positives (true positives plus false negatives) and reflects the model's ability to identify all relevant instances. The recall score was calculated as $\frac{TP}{TP+FN}$. The **F1 score** is the harmonic mean of precision and recall, providing a balanced measure that accounts for both false positives and false negatives. These metrics were computed using standard formulas and provide a comprehensive evaluation of the binary classification model's performance, particularly in situations where class distributions are balanced or when both false positives and false negatives are of equal concern.

For multiclass classification, we calculated weighted precision, recall, and F1 scores as $weighted\ score = \frac{\sum_k^K score_k \times t_k}{s}$. Here $K$ is the total number of classes (four for the multiclass classification task), $k$ is the index for a class, $t_k$ is the number of times class $k$ truly occurred, $s$ is the total number of samples, and $score_k$ is the evaluation metric score for class $k$. We additionally calculated the **Matthews Correlation Coefficient (MCC)** and **Cohen's Kappa** to provide a more nuanced assessment of model performance across all classes. The MCC is a robust metric that takes into account true and false positives and negatives across all classes, providing a balanced measure that remains effective even with imbalanced datasets. The MCC was calculated as $\frac{c \times s - \sum_k^K p_k \times t_k}{\sqrt{(s^2 - \sum_k^K p_k^2) \times (s^2 - \sum_k^K t_k^2)}}$. Here $p_k$ is the number of times class $k$ was predicted, and $c$ is the total number of samples correctly predicted. For multiclass classification, the minimum value of MCC will be somewhere between -1 and 0, and the maximum value will be +1, where +1 indicates perfect prediction, 0 indicates random prediction and a value close to -1 indicates complete disagreement between prediction and true class labels. **Cohen's Kappa** measures the agreement between predicted and actual labels while adjusting for the agreement that could occur by chance. It also ranges from -1 to +1, with values closer to +1 indicating strong agreement. Cohen's Kappa was calculated as $Kappa = \frac{p_o - p_e}{1 - p_e}$. Here $p_o = \frac{\sum_k^K C_{ii}}{s}$ is the observed agreement, calculated as the sum of the diagonal elements of the confusion matrix, and $p_e = \frac{\sum_k^K \left( \left( \sum_j^K C_{ij} \right) \left( \sum_j^K C_{ji} \right) \right)}{s^2}$ is the expected agreement by chance, where $C_{ij}$ is the element of the confusion matrix at row $i$ (representing true classes) and column $j$ (representing predicted classes). Both MCC and Cohen's Kappa offer insight into the overall performance of the classifier by considering all elements of the confusion matrix, making them particularly valuable for evaluating multiclass models where accuracy alone might be misleading due to class imbalance.

### S10. Hyperparameter tuning

We used 1,549 LNP formulations (Data14, 15, and 16, as shown in Table S4) for hyperparameter tuning of various ML algorithms with input features generated using RDKit descriptors. As shown in Figure S15 for the random forest classifier, the evaluation metrics were high, with small variance across the number of trees. For binary classification, the RF model with 70 decision trees (approximately one-third of the number of input features) performed the best. Based on these results, a value of 70 was used for the hyperparameter 'number of trees' for subsequent model training. As shown in Figure S16 for the gradient boosting classifier, the model performance improved noticeably as the number of boosting stages increased up to 200. Based on these observations, a value of 200 was used for the hyperparameter 'number of boosting stages' for subsequent model training. As shown in Figure S17 for the k nearest neighbors classifier, the model performance improved as up to five neighbors were included in the calculation, followed by a decline in performance with a further increase in the number of neighbors. Based on these observations, a value of 5 was used for the hyperparameter 'number of neighbors' for subsequent model training. As shown in Figure S18 for the logistic regression classifier, the difference in the accuracy of classification after 400 and 800 iterations was less than 1%. Based on these observations, a value of 400 was used for the hyperparameter 'maximum number of iterations' for subsequent model training.

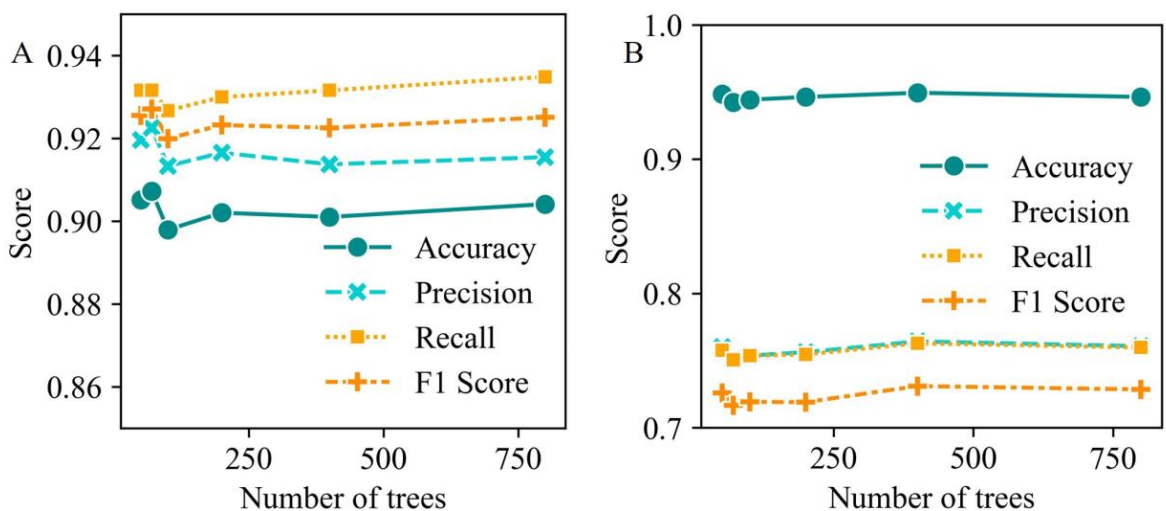

**Figure S15.** Performance of the random forest ML model for different values of the 'number of trees' parameter in (A) binary and (B) multiclass classification of LNPs based on activity.

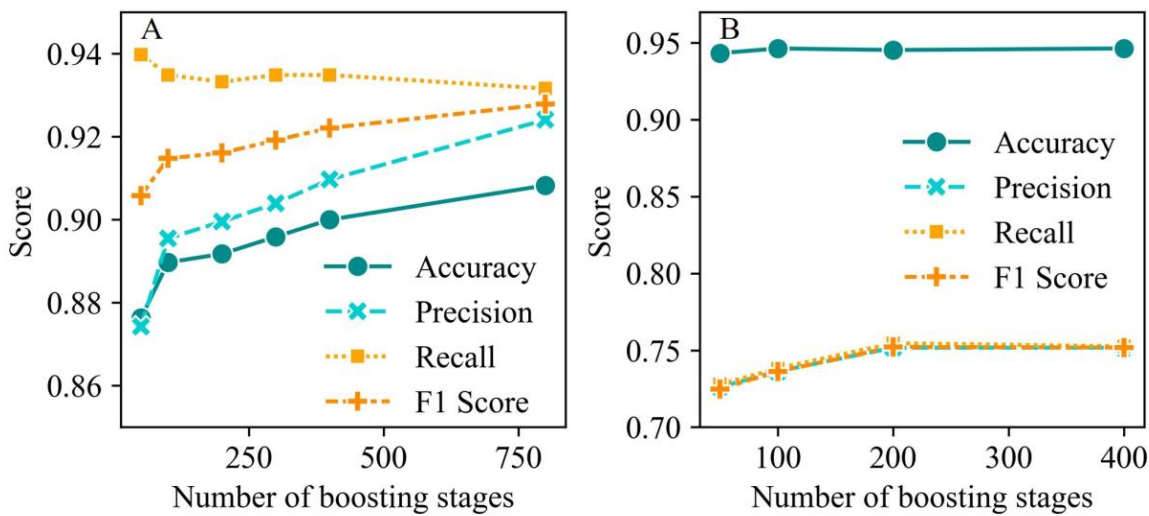

**Figure S16.** Performance of the gradient boosting ML model for different values of the 'number of boosting stages' parameter in (A) binary and (B) multiclass classification of LNPs based on activity.

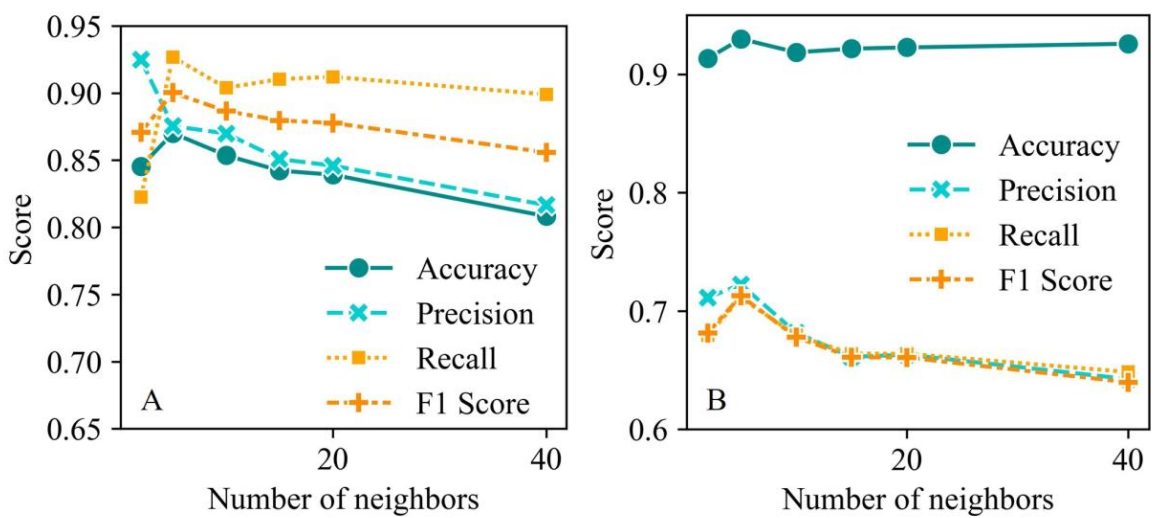

**Figure S17.** Performance of the k nearest neighbors ML model for different values of the 'number of neighbors' parameter in (A) binary and (B) multiclass classification of LNPs based on activity.

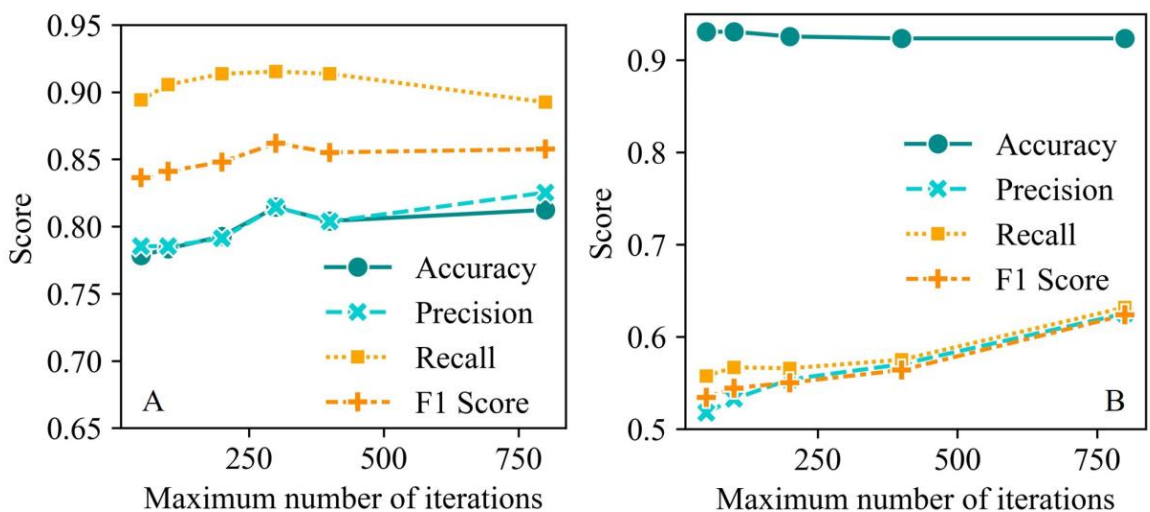

**Figure S18.** Performance of the logistic regression ML model for different values of the 'maximum number of iterations' parameter in (A) binary and (B) multiclass classification of LNPs based on activity.

## REFERENCES

(1) Liu, S.; Cheng, Q.; Wei, T.; Yu, X.; Johnson, L. T.; Farbiak, L.; Siegwart, D. J. Membrane-Destabilizing Ionizable Phospholipids for Organ-Selective MRNA Delivery and CRISPR–Cas Gene Editing. *Nat. Mater.* **2021**, *20*, 701–710.

(2) Lee, S. M.; Cheng, Q.; Yu, X.; Liu, S.; Johnson, L. T.; Siegwart, D. J. A Systematic Study of Unsaturation in Lipid Nanoparticles Leads to Improved MRNA Transfection In Vivo. *Angew. Chem. Intl. Ed.* **2021**, *60*, 5848–5853.

(3) Yu, X.; Liu, S.; Cheng, Q.; Wei, T.; Lee, S.; Zhang, D.; Siegwart, D. J. Lipid-Modified Aminoglycosides for MRNA Delivery to the Liver. *Adv. Healthc. Mater.* **2020**, *9*.